\documentclass[12pt,a4paper,conference]{article}

\usepackage{mathptmx} 
\usepackage{authblk} 

\usepackage[english]{babel}
\usepackage[utf8]{inputenc}
\usepackage[left=1.27cm,right=1.27cm,top=2cm,bottom=2cm]{geometry} 

\usepackage{graphicx} 
\usepackage[hang,small,justification=centering]{caption} 
\usepackage{float}
\usepackage[hidelinks]{hyperref}

\usepackage{tikz}
\usepackage[export]{adjustbox}
\usepackage{amsmath}
\usepackage{amssymb}
\usepackage{physics}
\usepackage{subcaption}
\usepackage{array}
\usepackage{etoolbox}

\newcommand{\rom}[1]{\textcolor{black}{#1}}

\newcommand{\fig}[1]{Fig.~\ref{#1}}
\newcommand{\eq}[1]{Eq.~(\ref{#1})}
\newcommand{\mypar}[1]{\vspace{0.3cm}\noindent\textbf{#1}}

\newcommand{\vz}{w} 

\newcommand{\az}{w_{,t}} 
\newcommand{\sx}{\eta_{,x}} 

\newcommand{\de}{\eta_{,t}} 
\newcommand{\dde}{\eta_{,tt}} 
\newcommand{\etaN}{\eta_0} 
\newcommand{\dEtaN}{\dot{\eta}_0} 

\newcommand{\cZ}{\check{Z}}

\newcommand{\cZb}{\check{Z}_b}

\newcommand{\nuz}{\mu_{_0}^{_\uparrow}}

\newcommand{\nub}{\mu_{b}^{_\uparrow}}
\newcommand{\etaP}{\eta_p} 
\newcommand{\etaB}{\eta_b} 
\newcommand{\dEtaB}{\dot{\eta}_b}
\newcommand{\ddEtaB}{\ddot{\eta}_b}
\newcommand{\fymax}{{F^{\perp}}_{\rm  max}}
\newcommand{\jy}{J^{\perp}}
\newcommand{\tip}{t_{\rm ip}}
\newcommand{\tex}{t_{\rm we}}
\newcommand{\tsep}{t_{\rm sep}}
\DeclareCaptionLabelFormat{nospace}{#1 #2}
\date{} 

\begin{document}

\makeatletter
\renewcommand{\section}{\@startsection {section}{1}{\z@}%
             {-3.5ex \@plus -1ex \@minus -.2ex}%
             {2.3ex \@plus.2ex}%
             {\large\bfseries}}
\renewcommand{\subsection}{\@startsection {subsection}{1}{\z@}%
             {-3.5ex \@plus -1ex \@minus -.2ex}%
             {2.3ex \@plus.2ex}%
             {}}
\renewcommand{\subsubsection}{\@startsection {subsubsection}{1}{\z@}%
             {-3.5ex \@plus -1ex \@minus -.2ex}%
             {2.3ex \@plus.2ex}%
             {}}

\def\@seccntformat#1{\@ifundefined{#1@cntformat}%
   {\csname the#1\endcsname\space}
   {\csname #1@cntformat\endcsname}}
\newcommand\section@cntformat{\thesection.\space\space \space \space}       

\makeatother

\renewcommand{\tablename}{Tableau}
\renewcommand{\tableautorefname}{Tableau}

\includegraphics[scale=1]{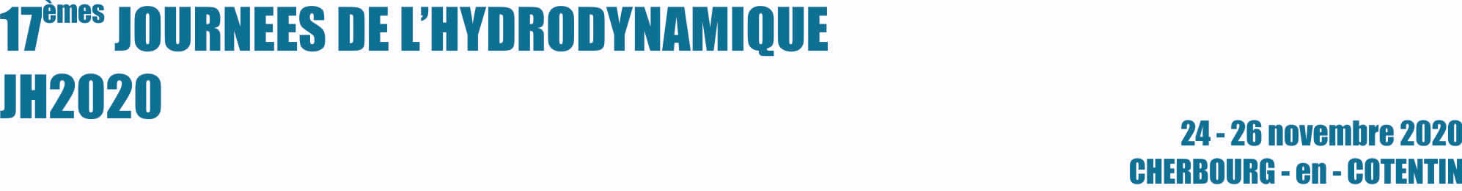}

\title{\large{\bf STOCHASTIC PREDICTION OF WAVE IMPACT KINEMATICS AND LOADS FOR SHIP APPENDAGES} \\ \textit{17$^{emes}$ JOURNEES DE L'HYDRODYNAMIQUE JH2020}}

\author[(1),*]{\bf R. Hasco\"et}
\author[(2)]{\bf M. Prevosto}
\author[(2)]{\bf N. Raillard}
\author[(1)]{\bf N. Jacques}
\author[(2)]{\bf A. Tassin}

\affil[(1)]{ENSTA Bretagne, CNRS UMR 6027, IRDL, 2 rue Fran\c{c}ois Verny, 29806 Brest Cedex 9, France}
\affil[(2)]{IFREMER -- LCSM, ZI Pointe du Diable, 29280 Plouzan\'{e} CS 10070, France}
\affil[*]{Corresponding author: romain.hascoet@ensta-bretagne.fr}

{\let\newpage\relax\maketitle}
\maketitle

\renewcommand{\abstractname}{Résumé\vspace{0.3cm}}
\begin{abstract}
\rom{Une approche stochastique est mise en oeuvre pour traiter la question 
d'une structure marine expos\'{e}e \`{a} des impacts de vagues.
L'\'{e}tude se concentre sur 
(i) la fr\'{e}quence moyenne des impacts de vagues 
et 
(ii) la distribution de probabilit\'{e} des variables cin\'{e}matiques associ\'{e}es \`{a} ces impacts.
Le champ de vagues est mod\'{e}lis\'{e} comme la r\'{e}alisation d'un processus Gaussien.
Les mouvements de tenue \`{a} la mer du corps consid\'{e}r\'{e} sont pris en compte dans l'analyse.
Le couplage de l'approche stochastique avec un mod\`{e}le d'impact hydrodynamique 
est illustr\'{e} sur le cas d'\'{e}tude d'un aileron expos\'{e} \`{a} des impacts de vagues.}
\end{abstract}

\renewcommand{\abstractname}{Summary\vspace{0.3cm}}
\begin{abstract}
\noindent 
A stochastic approach is implemented to address the problem of a marine structure exposed to water wave impacts.
The focus is on 
(i) the average frequency of wave impacts, 
and 
(ii) the related probability distribution of impact kinematic variables.
The wave field is assumed to be Gaussian.
The seakeeping motions of the considered body are taken into account in the analysis.
The coupling of the stochastic model with a water entry model is demonstrated through 
the case study of a foil exposed to wave impacts.
\end{abstract}

\section{Introduction}

When a ship sails on the sea, her appendages may be exposed to water wave impacts. 
The appendages may be below (e.g. fin stabilisers) or above (e.g. diving planes on a surfaced submarine) the mean sea level.
From an engineering standpoint, 
it may be important to know, for a given sea state, 
(i) the average frequency of wave impacts, and (ii) the probability distribution of resulting hydrodynamic loads.
Such a knowledge would be valuable as a guidance for the structural design of the appendage. 
It would offer the possibility to implement a design approach taking into account the expected extreme impacts 
(which could damage the structure by inducing stresses exceeding the material yield strength),
as well as repeated milder impacts (which could 
lead to structural failure through fatigue damage)

If the considered body is sufficiently small compared to water wave wavelengths 
-- which is usually the case for appendages --
the body geometry may be reduced to a single material point regarding the question of impact occurrence.
Then, an impact event can be modelled as the upcrossing of the material point by the sea free surface.
Thus, the problem becomes more tractable and
may be addressed by using the level crossing theory of stochastic processes,
based on the pioneering work of Rice (1944,1945) \cite{rice_1944, rice_1945}
and subsequent works (see for example \cite{lindgren_2012}, Chapter 8, and references therein).
When the motions of the waves and those of the ship are modelled at the first order, 
the randomness of related kinematic variables
can be modelled through Gaussian processes.
In this framework, the average impact frequency  
and the related joint probability distribution of kinematic variables can be analytically computed.
The distribution of kinematic variables may be transferred through an impact model
to compute the probability distribution of hydrodynamic loads.
If the impact model is 
computationally efficient
(e.g. based on a Wagner-type approach), 
the transfer may be achieved via a Monte Carlo method.
Helmers et al. (2012) \cite{helmers_2012} used such an approach to
investigate the probability distribution of impact loads 
for a wedge-shaped body exposed to irregular waves.
Conversely, if the impact model is numerically demanding (e.g. CFD simulations)
other approaches may be used, such as metamodels or reliability methods.

The present study investigates the stochastic framework briefly described above, 
taking into account the effect of seakeeping motions in the analysis.
Section \ref{sec_theoretical_framework} develops the theoretical framework 
to reach analytically the impact frequency 
and the related probability distribution of impact kinematic variables.
In Section \ref{sec_load}
the stochastic approach is coupled with a water entry model 
to compute the 
distribution of hydrodynamic loads 
on a foil exposed to wave impacts.


\begin{figure}[!b]
\begin{center}
\begin{tabular}{c}
        \includegraphics[width=0.62\textwidth]{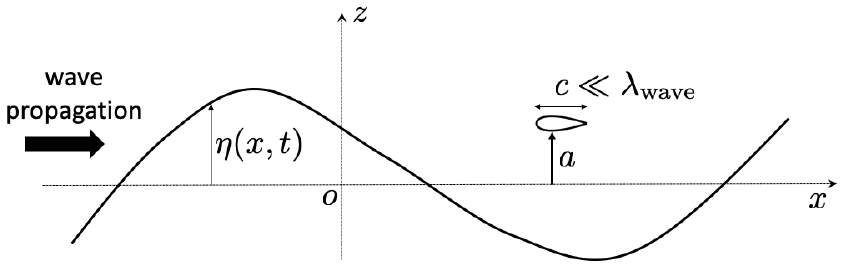}    
\end{tabular}
\end{center}
\vspace{-0.8cm}
\caption{
Illustration of the considered problem.
The sea state is modelled as a Gaussian process.
The body lies at a given average altitude $a$; it may have seakeeping motions. 
The characteristic size of the body (here sketched as a foil of chord $c$) 
is assumed to be much smaller than the water wave wavelengths, $\lambda_{\rm wave}$.
}
\vspace{-0.3cm}
\label{fig_sketch}
\end{figure}


\section{Theoretical framework: impact frequency and conditional distribution of kinematic variables}
\label{sec_theoretical_framework}

\subsection{Sea state modelled as a Gaussian process}

Throughout the present proceedings, 
sea states are considered to be unidirectional
and the water depth is assumed to be infinite.
These two assumptions are adopted for the sake of simplicity. 
However, the developments reported in the present document may be readily generalised to multi-directional sea states 
and finite water depth.
The problem is formulated in a coordinate system $Oxz$, 
which is attached to the reference frame of the mean flow 
(see Fig. \ref{fig_sketch} for an illustration of the problem).
The wave field is modelled as the superposition of independent Airy waves which follow the linear wave dispersion relation, 
$k = \omega^2 / g$,
where $\omega$ is the angular frequency, $k$ the wave number and  $g$ the acceleration due to gravity. 
In the framework of the linear wave model, the free surface elevation $\eta(x,t)$, 
can be modelled as a Gaussian process,  which is stationary in time and homogeneous in space.
It is fully characterised (in a probabilistic sense) by its variance density spectrum $S(\omega)$ 
(commonly referred to as the wave spectrum) and its mean (which is zero in the present case).

\subsection{Considered variables}

In the present 
study, the kinematic variables, whose distributions given up-crossing are investigated, are\\
\indent(i) $\vz(x,t)$, the vertical velocity of the fluid at $z=0$, which is equal to\footnote{
In the present document, $Q_{,v}$ denotes the derivative of the function $Q$ with respect to the variable $v$.
} $\de(x,t)$ to the first order, \\
\indent(ii) $\az(x,t)$, the vertical acceleration of the fluid at $z=0$, which is equal to $\dde(x,t)$ to the first order, \\
\indent(iii) $\eta_{,x}(x,t)$, the slope of the free surface.\\
The vertical velocity and acceleration, expressed at $z=0$, 
are considered as a direct proxy for the kinematics at the free surface.
When seakeeping motions are taken into account (see \S\ref{subsec_seakeeping}), 
the following additional kinematic variables are considered: \\
\indent(iv) $\dEtaB(t)$, the vertical velocity of the fluid measured in the reference frame of the material point, \\
\indent(v) $\ddEtaB(t)$, the vertical acceleration of the fluid measured in the reference frame of the material point, \\
\indent(vi) $\delta \theta(t)$, the pitch motion of the floating platform.\\
\noindent The variables $\dEtaB$, $\ddEtaB$ and $\delta \theta$ 
have been added as being relevant to compute the hydrodynamic loads related to the water entry event.
Including other kinematic variables in the stochastic analysis 
-- such as the horizontal velocity of the fluid, 
the horizontal acceleration of the fluid or the local curvature of the free surface --
would be straightforward.

\subsection{Body at rest}
\label{subsec_body_rest}

When the body is assumed to be at rest at a location $(x=x_0,z=a)$,
an impact event stands for the up-crossing of the level $a$ by the stochastic process  
\begin{equation}
\label{eq_eta}
\etaN(t) = \eta(x_0,t) \, ,
\end{equation}
whose time derivative is given by
\begin{equation}
\label{eq_eta}
\dEtaN(t) = \de(x_0,t) \, .
\end{equation}
Gathering the variables of interest, one may define the random vector 
\begin{equation}
Z = \left[ 
\begin{array}{l}
\eta = \etaN  \\
\dde  \\
 \de = \dEtaN  \\
 \sx  
 \end{array}
\right] \, .
\end{equation}

\mypar{Probability distribution given impact.}
$Z$ is a zero-mean Gaussian vector, whose covariance matrix $\Sigma_Z$ 
can be expressed as 
\begin{equation}
\label{eq_cov_mat_general}
\displaystyle \left[ \Sigma_Z \right]_{k,l} = \int_0^{+\infty} {\rm d} \omega \ {\rm Re}\left\{ \mathcal{H}_k(\omega) \bar{\mathcal{H}}_l(\omega)  \right\} S(\omega)  \, ,
\end{equation}
where $\mathcal{H}_k(\omega)$ is the linear transfer function whose input and output are respectively 
the sea surface elevation $\eta$ and the $k$-th variable of the random vector $Z$
and $\bar{\mathcal{H}}_l(\omega)$ denotes the complex conjugate of $\mathcal{H}_l(\omega)$.
In the present case, the linear wave theory yields the following transfer functions:
\begin{equation}
\label{eq_transFunc_0}
\begin{array}{l}
\mathcal{H}_\eta (\omega) = 1\\
\mathcal{H}_{\dde} (\omega) = - \omega^2 \\
\mathcal{H}_{\de} (\omega) = i \omega \\
\mathcal{H}_{\eta_{,x}} (\omega) = - i \omega^2/g \, ,
\end{array}
\end{equation}
where $i$ is the imaginary unit.
As these transfer functions are either real or imaginary, 
$Z$ can be split into two independent Gaussian random vectors,
\begin{equation}
\label{eq_vectX}
X = \left[ 
\begin{array}{l}
\eta  = \etaN \\
\dde 
\end{array} 
\right] 
\text{ and \ }
Y = \left[ 
\begin{array}{l}
 \de = \dEtaN \\
 \sx 
\end{array} 
\right]
\, ,
\end{equation}
whose covariance matrices are given by
\begin{equation}
\label{eq_sigmaX}
\Sigma_X = 
\left[
\begin{array}{cc}
m_0  & -m_2 \\
-m_2 & m_4 \\
\end{array}
\right]
\text{ and \ }
\Sigma_Y = 
\left[
\begin{array}{cc}
m_2 & -m_3/g \\
-m_3/g & m_4/g^2  \\
\end{array}
\right] \, .
\end{equation}
In \eq{eq_sigmaX}, $m_p$ denotes the $p$-th moment of the wave spectrum:
\begin{equation}
m_p = \int_0^{+\infty} {\rm d} \omega \ \omega^p S(\omega)  \, .
\end{equation}
In terms of probability density function, the independence of $X$ and $Y$ translates into
\begin{equation}
f_Z(\eta, \dde, \de, \sx  ) = f_X(\eta, \dde) \times f_Y( \de, \sx  ) \, ,
\end{equation}
where $f_Z$, $f_X$, $f_Y,$ are the probability density functions of the random vectors $Z$, $X$, $Y$, respectively. 
Let $\cZ$ denote the random vector containing the variables of $Z$, 
except for $\eta$.
The conditional density function of $\cZ$,  given that $\etaN$ is up-crossing the level $a$, can be written as
(see for example \cite{lindgren_2012}):
\begin{equation} 
\label{eq_fZ_rest}
f_{\cZ|\etaN\uparrow a}(\dde, \de, \sx) = \frac{\de \cdot f_{\cZ|\eta= a}(\dde, \de, \sx)}
{\displaystyle \int_0^{+\infty} \! \! \! \! {\rm d} \xi \ \xi  f_{\de|\eta= a}(\xi)} , \ {\rm with} \ \de > 0 \, ,
\end{equation} 
where $f_{\cZ|\eta= a}$ and $f_{\de|\eta= a}$ are the conditional density functions of $\cZ$ and $\de$, given $\eta=a$.
Taking advantage of the independence of the vectors $X$ and $Y$, 
\eq{eq_fZ_rest} can be further expressed as follows:
\begin{equation} 
\label{eq_fZ_2}
f_{\cZ|\etaN\uparrow a}  (\dde, \de, \sx)
= f_{\dde|\eta=a}(\dde)  \times
\sqrt{\frac{2\pi}{m_2}}  \de  f_{Y}(\de, \sx)
\, , \ {\rm with} \ \de > 0 \, .
\end{equation} 

\mypar{Impact frequency.}
The average up-crossing frequency, i.e. the impact frequency, at a given altitude $a$,
is given by Rice's formula \cite{rice_1945}:
\begin{equation}
\label{eq_nuz_general}
\nuz(a) = \int_0^{+\infty}  \dEtaN f_{\etaN, \dEtaN} (a, \dEtaN) \ {\rm d} \dEtaN \, .
\end{equation}
As $\etaN$ and $\dEtaN$ are independent normal variables,
\eq{eq_nuz_general} can be readily further expressed as:
\begin{equation}
\label{eq_fip_rest}
\nuz(a) = \frac{1}{2\pi} \sqrt{\frac{m_2}{m_0}} \exp \left( - \frac{a^2}{2 m_0} \right) \, .
\end{equation}
%


\def\scaleF{0.3}

\begin{figure}[h!]
\begin{center}
\begin{tabular}{ccc}
        \includegraphics[width=\scaleF\textwidth]{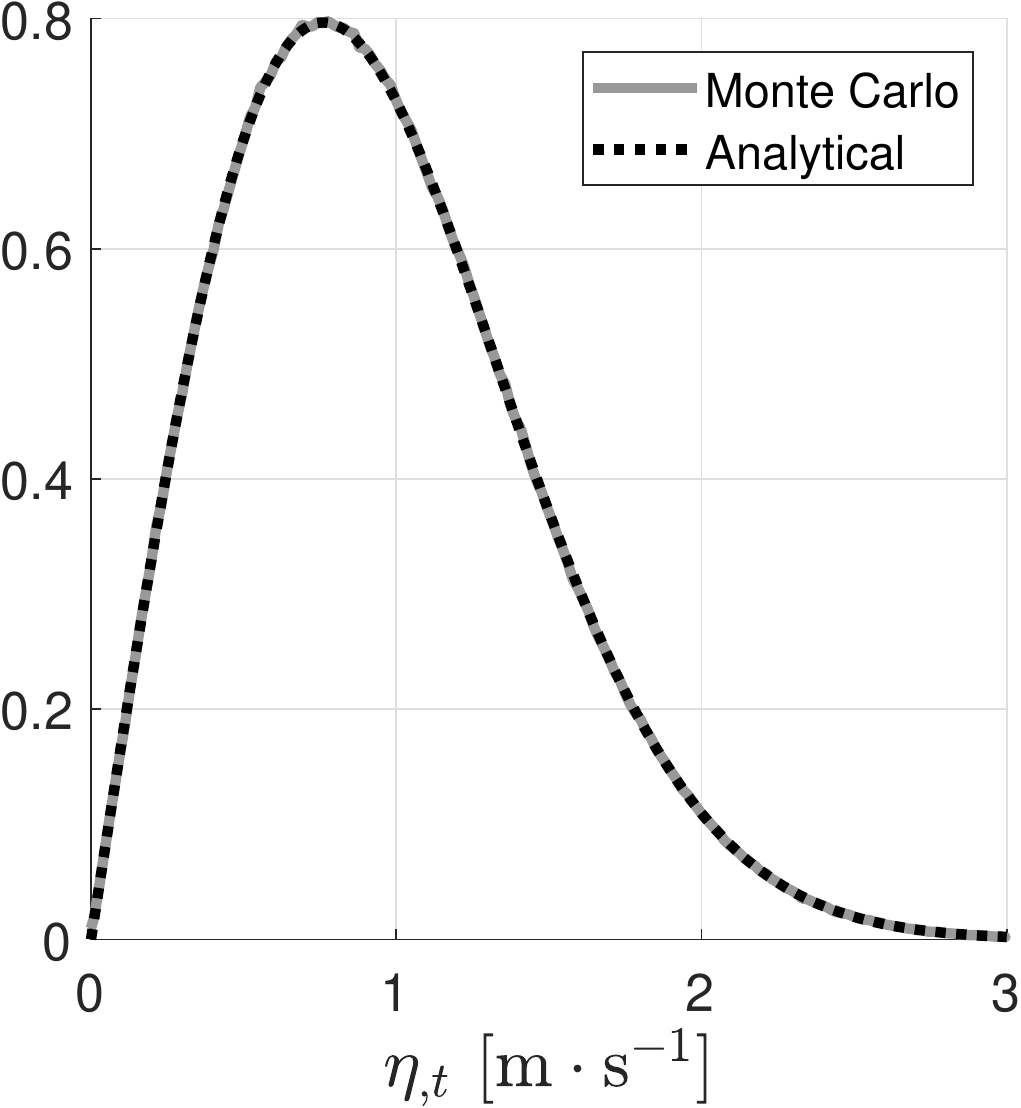} & 
        \includegraphics[width=\scaleF\textwidth]{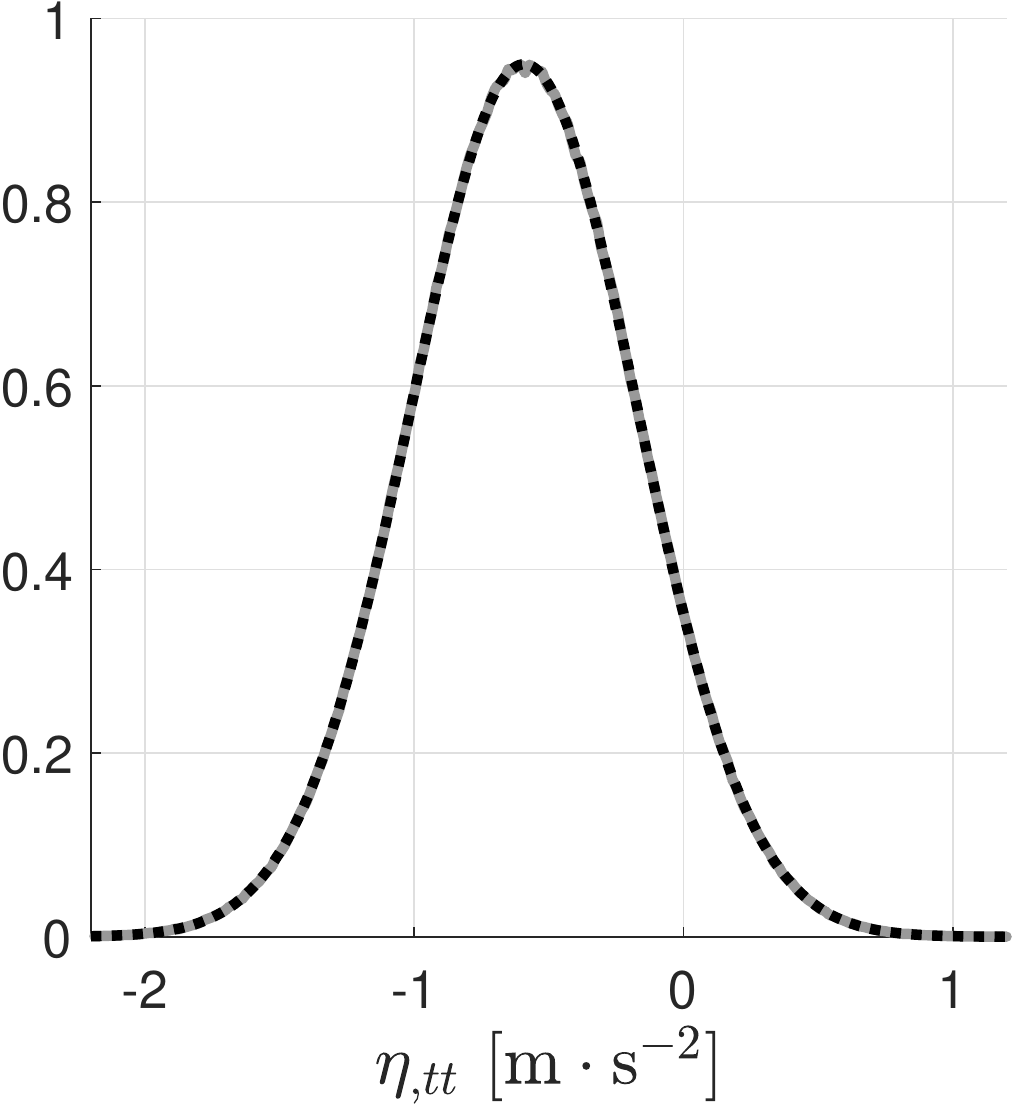} &
        \includegraphics[width=\scaleF\textwidth]{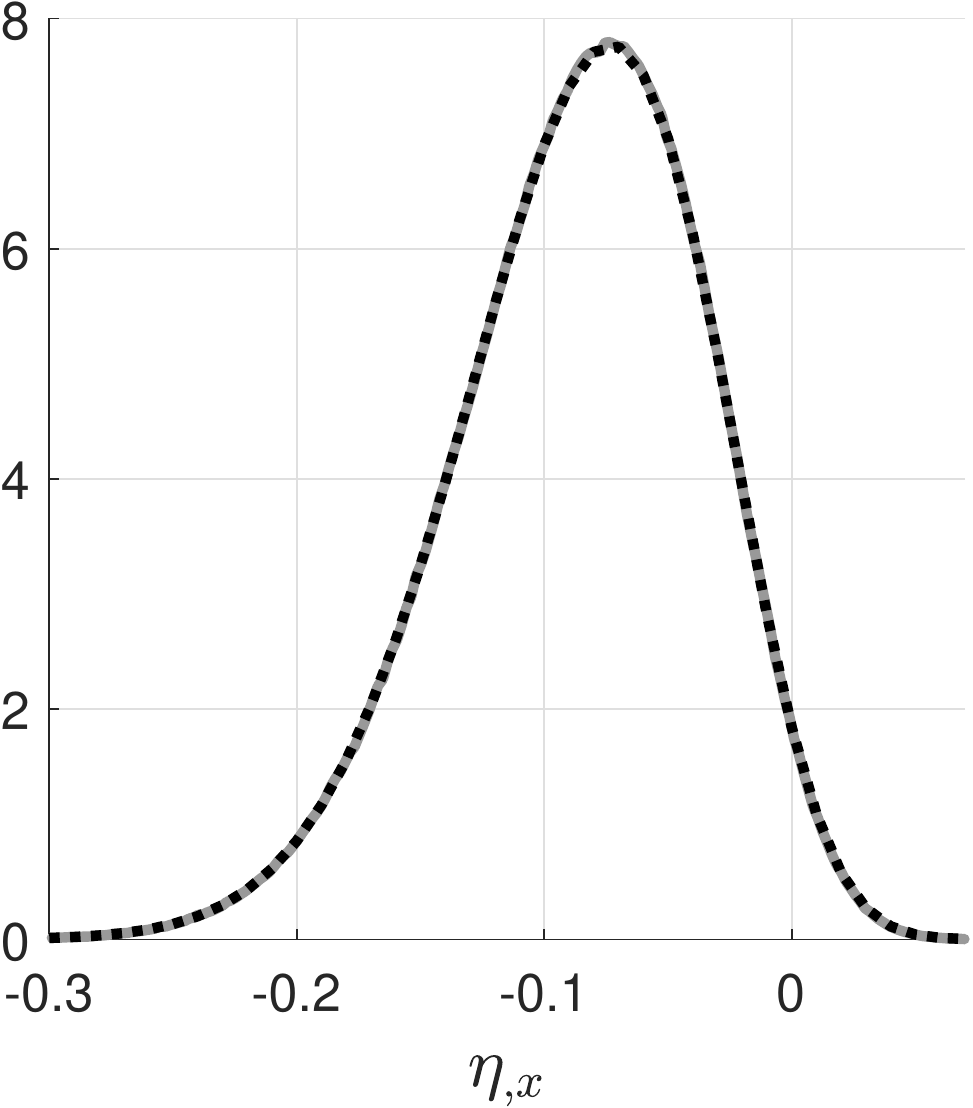}      
\end{tabular}
\end{center}
\caption{
Univariate probability density functions of random kinematic variables, given impact, 
for a body at rest.
The wave spectrum is assumed to have a JONSWAP shape with a peak-shape parameter $\gamma=3.3$, 
a significant wave height $H_s = 4$ m, and a peak period $T_p = 10$ s.
The material point is fixed at an altitude $a=1$ m.
The dashed black lines show the results obtained by integrating the analytical multivariate density function given in \eq{eq_fZ_2}.
The solid grey lines show the numerical results obtained from Monte Carlo simulations 
(multiple realisations of the sea state).
}
\label{fig_example_rest}
\end{figure}


\mypar{Illustrative example.}
As an illustrative example, Fig. \ref{fig_example_rest} shows the univariate probability density functions of kinematic variables,
obtained for a sea state with $H_s = 4$ m, $T_p = 10$ s, and a material point fixed at an altitude $a=1$ m.
For testing purpose, the probability density functions have been computed by using two different methods: 
(i) analytical calculation by successive integrations of the multivariate density function given in \eq{eq_fZ_2},
(ii) Monte Carlo simulation of many realisations of the stochastic process, based on the random-phase/amplitude model 
(see e.g. \cite{holthuijsen_2007}).
Both methods show an excellent agreement on the conditional density function of kinematic variables. 
It was also checked that the up-crossing frequency obtained from the simulations 
is in good agreement with the analytical expression given in \eq{eq_fip_rest}.
The analytical integration of \eq{eq_fZ_2}
-- not detailed here -- shows that: 
(i) the conditional distribution of $\de$ is of Rayleigh-type with a mode equal to $\sqrt{m_2}$, 
(ii) the conditional distribution of $\dde$ is normal,
(iii) the conditional distribution of $\sx$ is the convolution product of a Rayleigh distribution with a normal distribution.
A modification of the chosen altitude $a$ would change the results only through a shift in the conditional mean of $\dde$.

\subsection{Body with seakeeping motions}
\label{subsec_seakeeping}

Accounting for seakeeping motions in the present stochastic approach, 
presents no specific difficulty, as long as the seakeeping motions are linearly modelled.
Waves are assumed to be unidirectional, approaching the floating platform 
from an angle $\psi$.
As an illustrative example,
let us consider the case of a 
ship which is symmetric about its longitudinal centreplane,
and is standing
in head seas ($\cos \psi = -1$) or following seas ($\cos \psi = 1$),
with no forward speed.
Then, the seakeeping motions to be considered are:
(i) the heave motion, $\delta z(t)$,
(ii) the surge motion,  $\delta x(t)$,
(iii) the pitch motion,  $\delta \theta (t)$.
The coordinates of the material point, in the frame of the mean flow,
are given by (assuming $\delta \theta$ to be small)
\begin{equation}
\label{eq_sk}
\begin{array}{l}
x(t) \simeq \left[ x_p + z_p \delta \theta(t) + \delta x(t) \right] \cos \psi + x_0 \\
z(t) \simeq a - x_p \delta \theta(t) + \delta z(t) ,
\end{array}
\end{equation}
where $(x_p, z_p)$ is the position of the material point in the frame used to compute 
the Response Amplitude Operators (RAOs) of the ship, 
and $x_0$ is the average longitudinal position of the origin of this frame.
The $x_p$-axis is oriented toward the bow, the $z_p$-axis is oriented upwards,
and the ship is nosed down for $\delta \theta > 0$.
$a$ is the altitude of the material point when the ship is at rest (hydrostatic equilibrium).
If the longitudinal seakeeping motion is small compared to the water wave wavelengths
(consistent with the linearity assumption), i.e. if
\begin{equation}
k [ z_p \delta \theta (t) + \delta x (t) ] \ll 1 \,  ,
\end{equation}
the seakeeping motion along the $x$-axis can be ignored in the stochastic problem.\footnote{
In the opposite case, the sea surface elevation at the location of the material point, would depend on the seakeeping response.
Then, the problem would become nonlinear, making analytical developments much less tractable.
}
In this framework, an impact event stands for the up-crossing of the stochastic process $z(t)$ 
by the stochastic process 
\begin{equation}
\etaP (t) = \eta(x_0 + x_p \cos \psi, t) \, ,
\end{equation}
which is the free surface elevation at the location of the material point.
By defining the stochastic process
\begin{equation}
\label{eq_zetab}
\etaB(t) = \etaP (t) + x_p \delta \theta(t) - \delta z(t),
\end{equation}
an impact event is also equivalent to the up-crossing of the level $a$ by the Gaussian process $\etaB(t)$.
To compute the distribution of kinematic variables, given impact, 
the following Gaussian vector is considered:
\begin{equation}
Z_b = \left[ 
\begin{array}{c}
\dde  \\
 \de \\
  \sx  \\
  \delta \theta \\
  \ddEtaB \\
 \etaB \\
 \dEtaB
 \end{array}
\right] \, .
\end{equation}
The variables $\etaB$ and $\dEtaB$ need to be considered 
since the up-crossing 
condition is checked for the stochastic process $\etaB(t)$.
The variables $\ddEtaB$ and $\delta \theta$ have been also included in the considered Gaussian vector, 
as being relevant (together with $\dEtaB$) to compute slamming loads.
The transfer function of $\etaB$ (whose input is $\eta = \etaP$), is given by
\begin{equation}
\label{eq_tf_zetab}
\mathcal{H}_{\etaB} = 1 + \exp\left\{ i k x_p \cos \psi \right\} (x_p \mathcal{H}_{\theta} - \mathcal{H}_{z}) \, ,
\end{equation}
where $\mathcal{H}_{\theta}$ is the pitch RAO and $\mathcal{H}_{z}$ is the heave RAO.
The phase-shift term, $\exp\left\{ i  k x_p \cos \psi \right\}$, accounts for the fact that the free surface elevation 
is monitored at the location of the material point (and not at the origin of the frame used to compute RAOs).
The transfer functions of the time derivatives $\dEtaB$ and $\ddEtaB$ are given by
\begin{equation}
\begin{array}{l}
\mathcal{H}_{\dEtaB}(\omega) = i \omega \mathcal{H}_{\etaB}(\omega) \\
\mathcal{H}_{\ddEtaB}(\omega) = - \omega^2 \mathcal{H}_{\etaB}(\omega) \, .
\end{array}
\end{equation}

\mypar{Probability distribution given impact.}
\textit{A priori}, the transfer functions $\mathcal{H}_{\etaB}$ and $\mathcal{H}_{\dEtaB}$ are neither purely real 
nor purely imaginary. 
As a result, contrary to the approach adopted in \S\ref{subsec_body_rest}.
the variables of the vector $Z_b$ cannot be divided into two independent Gaussian vectors.
The covariance matrix of $Z_b$ may be computed by an expression of the form of \eq{eq_cov_mat_general}.
Then, the probability density function of $\cZb$ 
(vector collecting the variables of $Z_b$, except for $\etaB$), given impact, can be expressed as
\begin{equation} 
\label{eq_fZ_b}
f_{\cZb|\etaB\uparrow a} = \frac{\sqrt{2\pi}}{{\sigma_{\dEtaB}}} \dEtaB \times f_{\cZb|\etaB=a} 
(\dde, \de, \sx, \delta \theta, \ddEtaB, \dEtaB)
\, , \ \dEtaB > 0 \, ,
\end{equation}
where $f_{\cZb|\etaB=a}$ is the conditional probability density function of the vector $\cZb$, given $\etaB=a$,
and ${\sigma_{\dEtaB}}$ is the non-conditional standard deviation of $\dEtaB$.

\mypar{Impact frequency.}
Similarly to \eq{eq_fip_rest},
the up-crossing frequency of the level $a$ by the sea surface elevation, i.e. the impact frequency, 
can be expressed as:
\begin{equation}
\label{eq_fip_sk}
\nub(a) = \frac{1}{2\pi}  \frac{\sigma_{\dEtaB}}{\sigma_{\etaB}} 
\exp \left( - \frac{a^2}{2 \sigma_{\etaB}^2} \right) \, ,
\end{equation}
where ${\sigma_{\etaB}}$ is the non-conditional standard deviation of $\etaB$.

\mypar{Illustrative example.}
As an illustrative example, let us consider the case of a ship
with no forward speed, in head seas.
The assumed heave and pitch RAOs (not reported here) 
correspond to a ship of length $L \simeq 100$ m.
The assumed sea state is the same as in \fig{fig_example_rest}.
By successively integrating \eq{eq_fZ_b}, 
the resulting univariate density functions, given up-crossing, are shown in \fig{fig_dists_sk}. 
The results are shown for three different material points attached to the ship,
$P_1$, $P_2$, $P_3$.
When the ship is at rest (hydrostatic equilibrium), 
the altitude of the three considered material points is $a=1$ m.
The longitudinal positions of $P_1$, $P_2$, $P_3$, in the frame of the ship,
are respectively 
$x_{p_{1}} = +50 $ m (bow),
$x_{p_{2}} = 0 $ (midship),
$x_{p_{3}} = -50 $ m (stern).

At the position $P_2$, the impact frequency, 
${\nub}@{P_2}  = 5.6 \cdot 10^{-2} \ {\rm Hz}$, 
is smaller compared to the case where no seakeeping motion is considered
($\nuz = 7.3 \cdot 10^{-2} \ {\rm Hz}$).
From a physical viewpoint, 
it is related to the fact that the material point tends to follow the elevation of the longest waves, 
due to the heave motion of the ship.
Conversely, for $P_1$ and $P_3$, the motion of the material point becomes out of phase with the local free surface elevation,
both because of the phase shift introduced by the longitudinal gap 
between the considered material point and the centre of the ship,
and because of the pitch motion.
This results in a larger impact frequency at positions $P_1$ and $P_3$, 
${\nub}@{P_1} = {\nub}@{P_3} = 7.9 \cdot 10^{-2} \ {\rm Hz}$.

The random variable $\dEtaB$, given up-crossing, follows a Rayleigh distribution 
with a mode equal to $\sigma_{\dEtaB}$.
The variable $\ddEtaB$, given up-crossing, follows a normal distribution.
When the material point has seakeeping motions, 
$\de$ given up-crossing is not Rayleigh-distributed anymore and negative values are allowed. 
Indeed the moving material point may enter the water domain, 
even though the free surface elevation, 
$\eta$, 
is locally decreasing.
Instead, the conditional distribution of $\de$, given up-crossing, 
results from the convolution product of a Rayleigh distribution with a normal distribution 
(calculation not detailed here).
Regarding the other variables, $\dde$, $\sx$, $\delta \theta$, 
they also follow distributions which can be expressed as the convolution product of a Rayleigh distribution with a 
normal distribution.


\def\scaleF{0.3}
\def\vertAnnot{1.95}
\def\horAnnot{-1.4}

\def\prefix{Hs_4_Tp_10_gamma_3p3_trunc_0p01_eta_1_Vs_0_SK_}

\begin{figure}[t]
\begin{center}
\begin{tabular}{ccc}
        \includegraphics[width=\scaleF\textwidth]{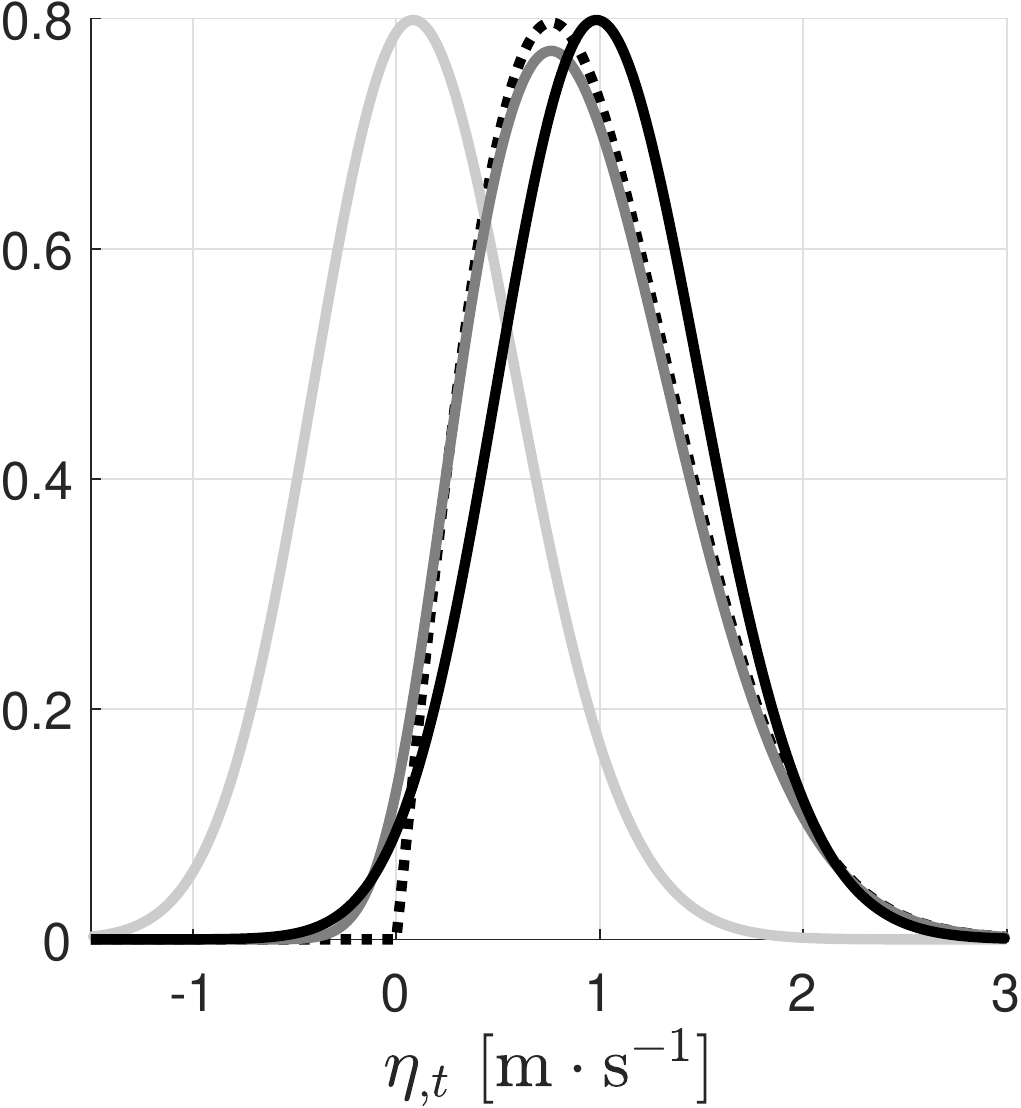} & 
        \includegraphics[width=\scaleF\textwidth]{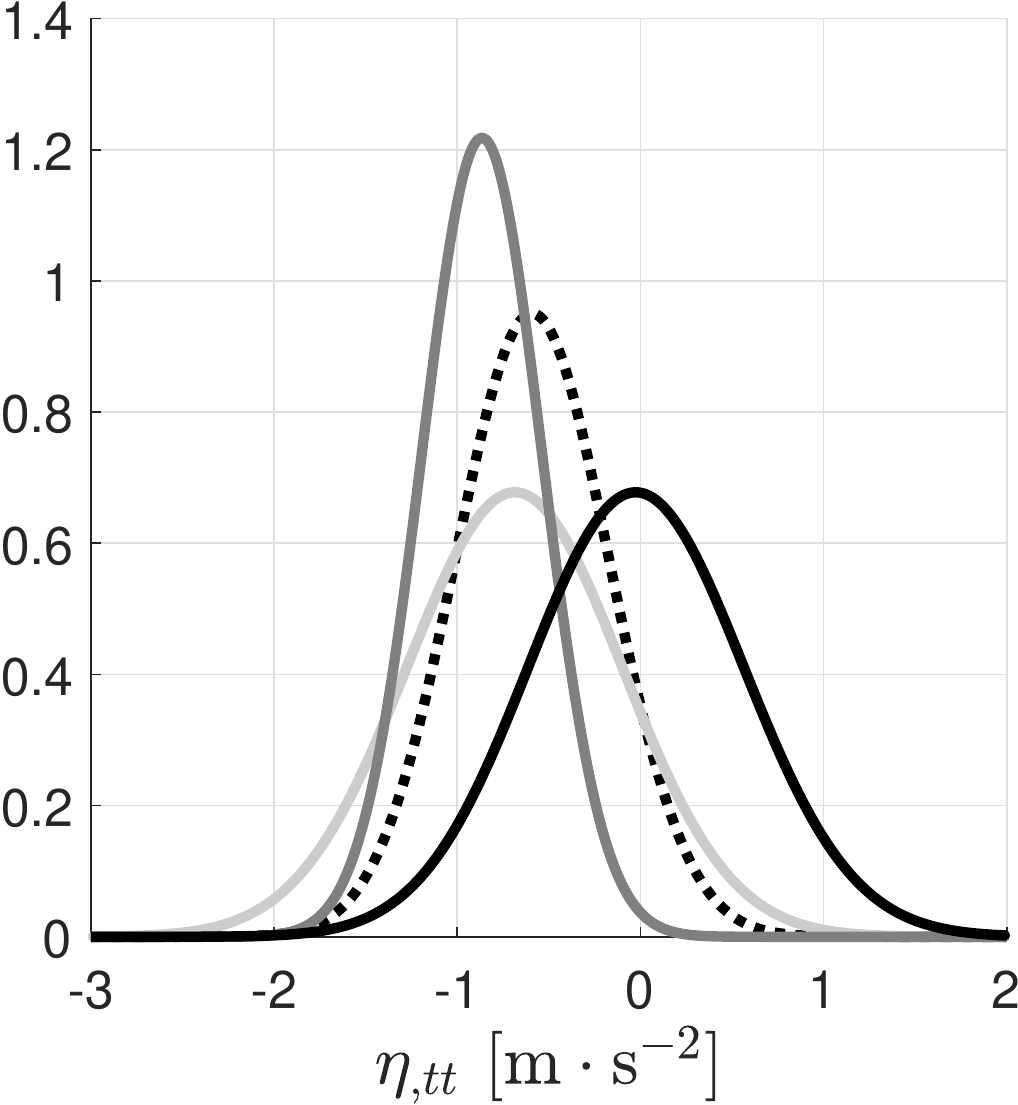} &
        \includegraphics[width=\scaleF\textwidth]{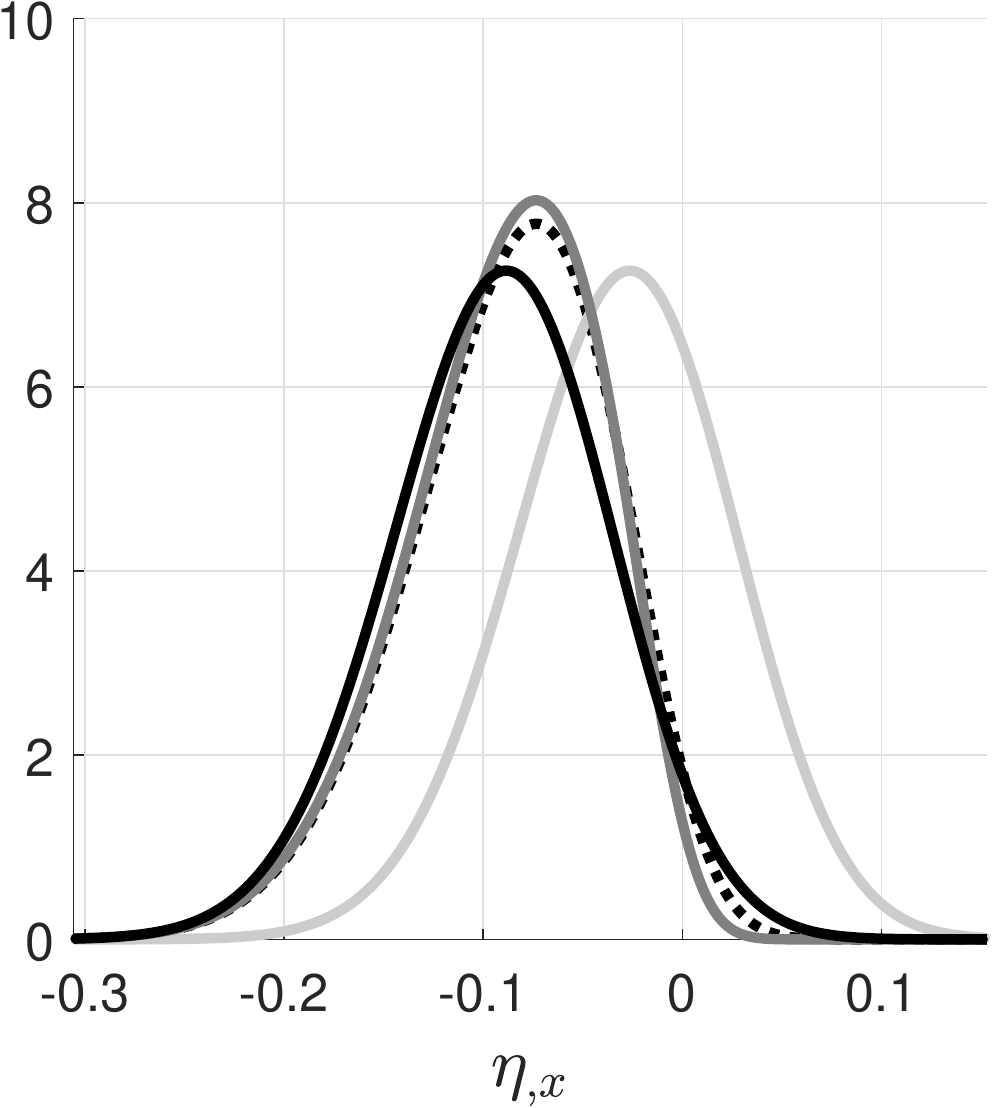}  \\
        \includegraphics[width=\scaleF\textwidth]{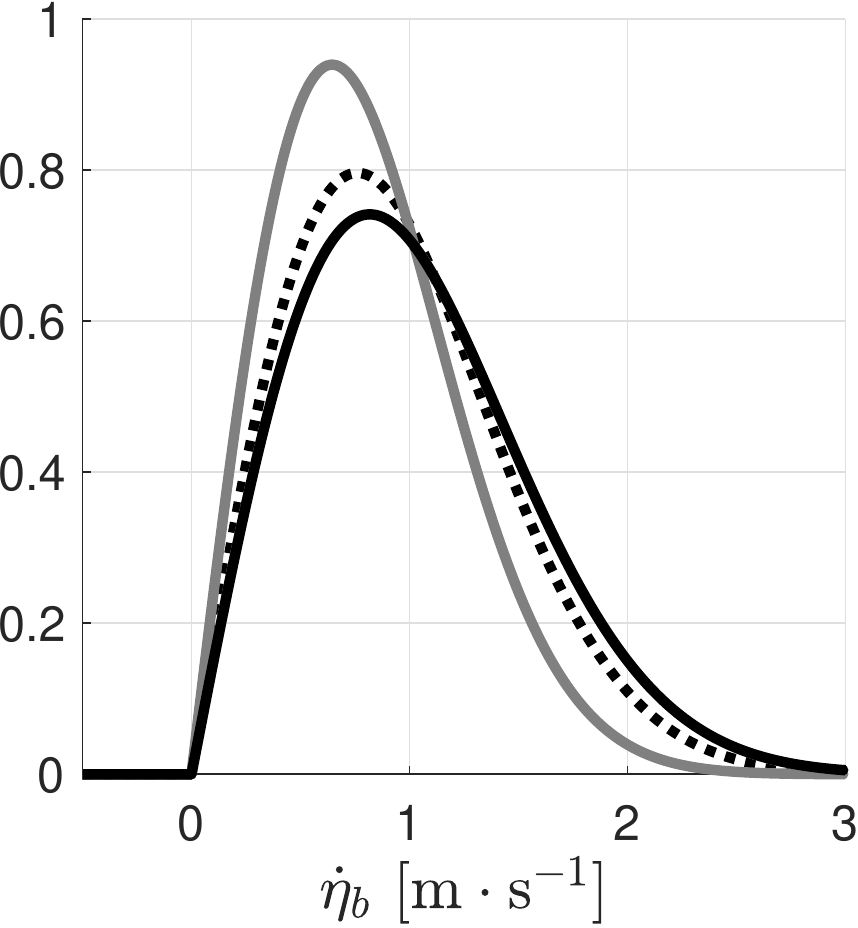} & 
        \includegraphics[width=\scaleF\textwidth]{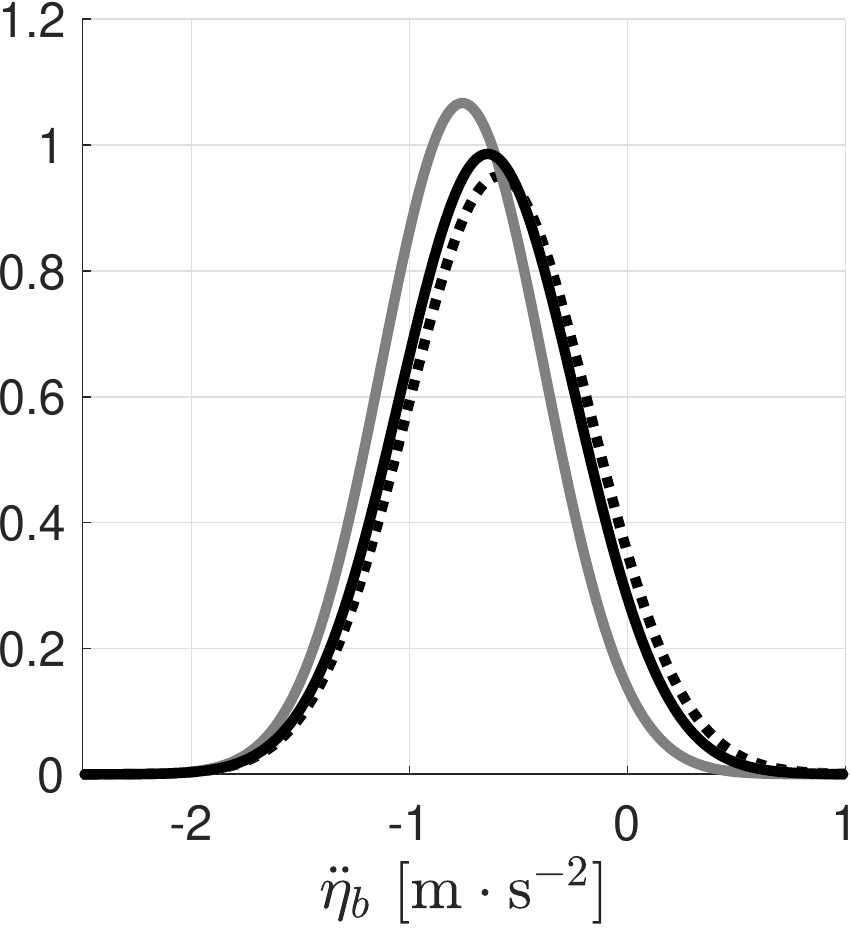} &
        \includegraphics[width=\scaleF\textwidth]{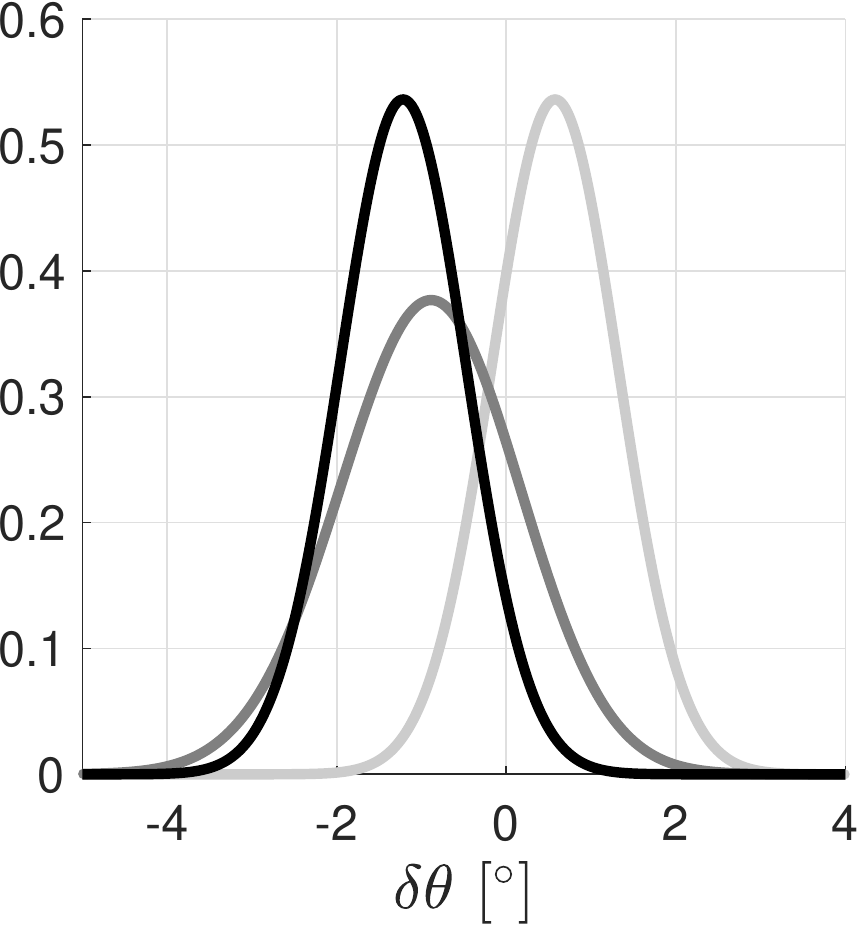}
\end{tabular}
\end{center}
\caption{
Probability density functions of random kinematic variables, 
given impact -- effect of seakeeping motions. 
Results obtained for a fixed platform (no seakeeping motion) are shown as dotted lines.
Results with seakeeping motions are shown for three different material points attached to the ship,
$P_1$ (bow, black curves), $P_2$ (midship, dark grey curves), $P_3$ (stern, light grey curves).
}
\label{fig_dists_sk}
\end{figure}

\section{Propagation of distributions through 
an impact model -- Illustrative example.}
\label{sec_load}

If the water entry model is fully analytical and simple enough, 
the transfer of the probability distribution of kinematic variables 
(Eqs. \ref{eq_fZ_2}-\ref{eq_fZ_b})
through the water entry model (to access hydrodynamic loads) may be performed analytically.
Such an approach has been implemented for instance in \cite{ochi_1973}
where the authors used a univariate quadratic model for the slamming pressure,
considering the fluid vertical velocity as the only relevant variable.
When the hydrodynamic model becomes more elaborate, with possibly multiple input variables,
a fully analytical transfer becomes more challenging, if not impossible 
(for instance if the impact model is based on CFD simulations). 
Then, the transfer may be performed by using alternative approaches, 
such as Monte Carlo sampling, metamodels, or reliability methods.

In the present study, 
the stochastic approach
has been coupled with the semi-analytical water entry model introduced by \cite{hascoet_2019b}. 
Combining Wagner's approach with the concept of ``Fictitious Body Continuation'' \cite{tassin_2014}, 
this water entry model can account for flow separation (when the fluid detaches from the edges of the body).
In the present case study, the considered body is the same as in \cite{hascoet_2019b}, 
i.e. a two-dimensional foil of section NACA 0028 (as represented in \fig{fig_sketch}).
The foil is assumed to be at rest (in the reference frame of the mean flow);
no seakeeping motion is considered.
The kinematic variables used as input of the impact model
are the relative fluid velocity $\vz = \de$, the relative fluid acceleration $\az=\dde$, 
and the orientation angle of the body with respect to the local free surface 
(this last variable is directly inferred from the local free surface slope, $\sx$).
The transfer of the distribution of kinematic variables through the impact model 
has been performed by use of a Monte Carlo sampling.
The wave slope $\sx$ is assumed to be fixed during the impact 
(due to a limitation of the water entry model), 
while the time evolution of $\vz$ and $\az$ is taken into account (see below).

\begin{figure}[t]
\begin{center}
\begin{tabular}{cc}
        \includegraphics[width=\scaleF\textwidth]{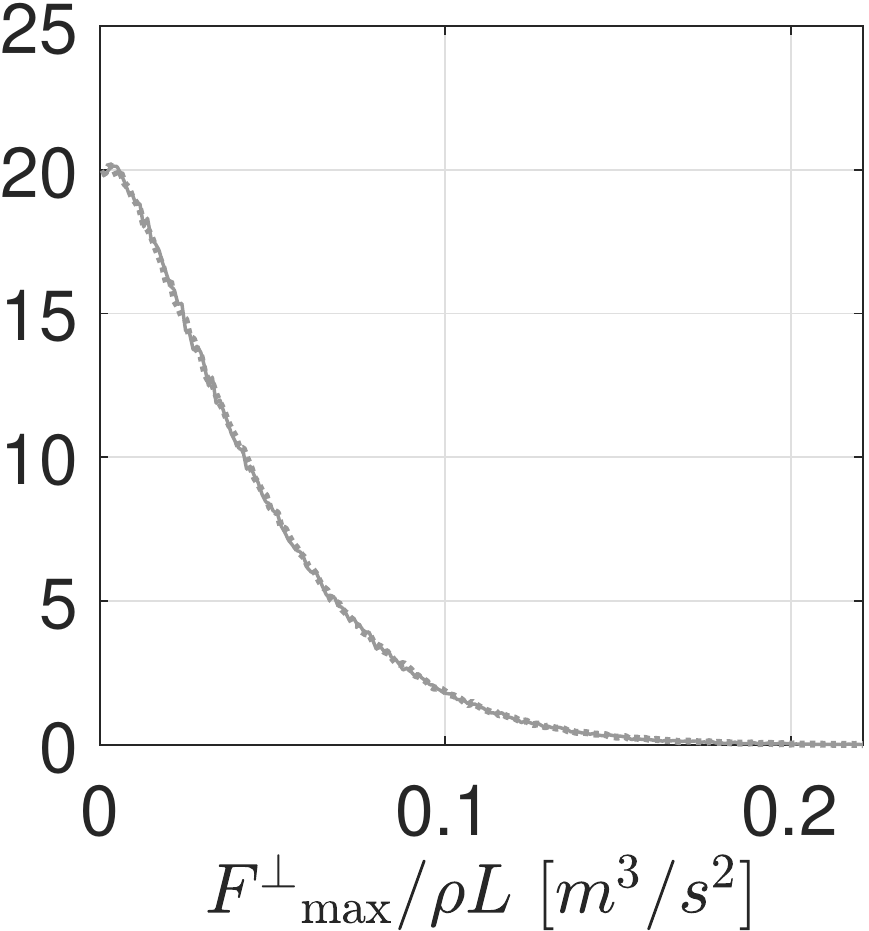} & 
        \includegraphics[width=\scaleF\textwidth]{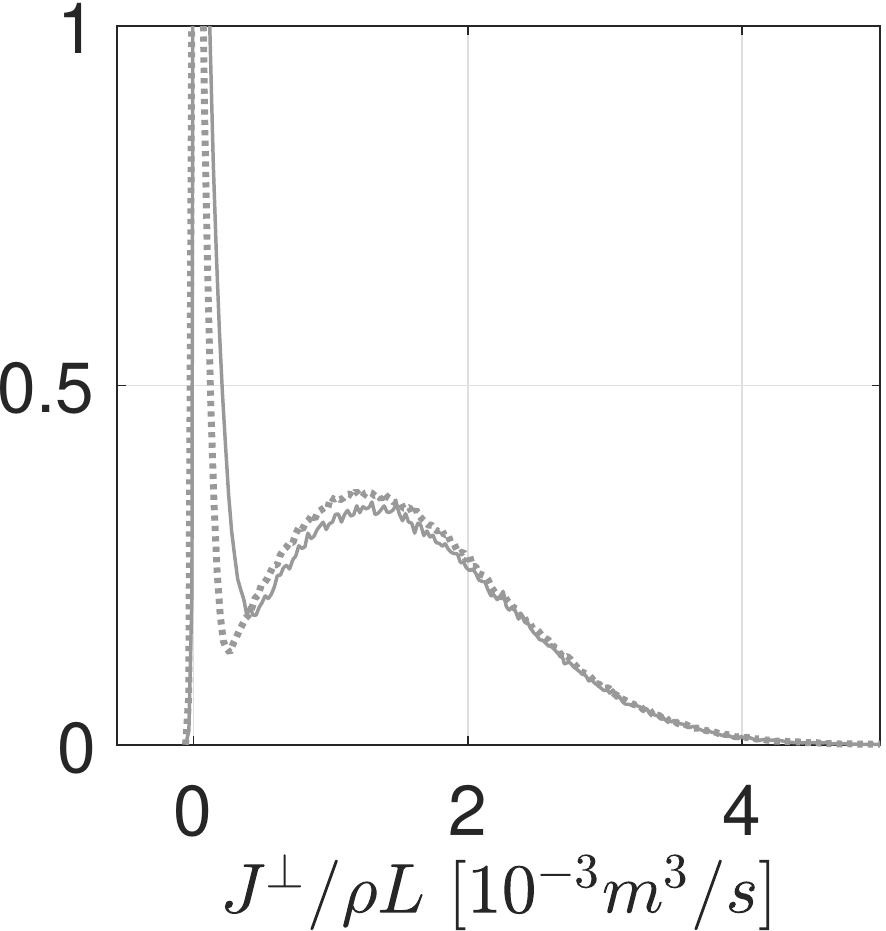} 
\end{tabular}
\end{center}
\caption{
Distribution of hydrodynamic loads. 
The assumed sea state has a significant wave height $H_s = 27$ cm, 
and a peak period $T_p = 3$ s (tank test configuration).
The body is a two-dimensional NACA 0028 foil with a chord $c=15$ cm. 
The foil is horizontal (relative to the mean sea level), 
located at an altitude $a=10$ cm (lowest point of the body), 
and facing the sea. 
The resulting impact frequency is $\nuz \simeq 0.13$ Hz.
The left and right panels show, respectively, the probability density functions of 
maximal impact force $\fymax$ 
and impulsion $\jy$ (divided by the fluid density, $\rho$, and the span of the foil, $L$). 
The results obtained from approaches (a) and (b), described in Section \ref{sec_load},
are respectively shown as dashed lines and solid lines.
In the left panel, both curves are almost superimposed.
The density function of $\jy$ is bimodal,
which is due to the dual definition of the impact duration, $\tip$ (see the body of the text).
The peak of the first mode (corresponding to mild impacts, with $\tip=\tex$) is cropped in the present figure.
}
\label{fig_transfer_hydro}
\end{figure}

As a sample of our results, \fig{fig_transfer_hydro} shows the probability density functions of two quantities: 
(i) $\fymax$, the maximum value (during a given impact) of ${F^{\perp}}$,
which is defined as the hydrodynamic force 
(per unit span of the foil)
component normal to the local free surface;
(ii) 
$\jy$, the time-integral
of ${F^{\perp}}$ (i.e. impulse) over the duration of the impact, $\tip$.
The duration of an impact event is defined as follows: 
if the water entry stops (i.e. the free-surface elevation starts decreasing, initiating a water exit phase) 
before the flow has detached from both sides of the body, then $\tip = \tex$,  
where $\tex$ is the duration of the water entry stage.
Otherwise, the considered impact duration correspond to the time, $\tsep$, when the flow has detached from both sides of the body 
(after flow separation, hydrodynamic loads start decreasing rapidly, see \cite{hascoet_2019b}).
Two different approaches were used to estimate hydrodynamic loads:
(a) approximate approach
assuming that the flow acceleration is constant during the impact
(hence only the initial values of $\vz$ and $\az$ need to be drawn),
(b) Monte Carlo simulation of the stochastic evolution of kinematic variables after up-crossing (this option is numerically more demanding).
Both methods give close results, which shows that the approximate approach (a) 
can provide satisfactory predictions
if the body is sufficiently small compared to wave dimensions 
(wave height and wavelength), as it is the case here.

\section*{Acknowledgements}
\small
The present work was supported by the French National Agency for Research (ANR) 
and the French Government Defense procurement and technology agency  (DGA) [ANR-17-ASTR-0026 APPHY];
the related research project involves SIREHNA, Bureau Veritas, IFREMER, and ENSTA Bretagne.

\bibliographystyle{ieeetr}
\bibliography{mybibfile}

\end{document}